\documentstyle{article}
\newcommand{\vecb}[1]{{\bf #1}}  

\newcommand{\pder}[2]{\frac{\partial {#1}}{\partial {#2}}}
\newcommand{\pderder}[2]{\frac{\partial^2 {#1}}{\partial {#2}^2}}

\newcommand{\reals}{\mbox{$I\!\!R$}}
\newcommand{\cpr}{\mbox{$c'$}}
\newcommand{\avc}{\mbox{$\overline{c}$}}

\newcommand{\eff}{{\mbox{\scriptsize eff}}}
\newcommand{\ReP}{\Re}
\newcommand{\cM}{{\cal M}}
\newcommand{\Ord}[1]{{\cal O}\left(#1\right)}
\newcommand{\xv}{{\vecb x}}
\newcommand{\yv}{{\vecb y}}
\newcommand{\zv}{{\vecb z}}
\newcommand{\hv}{{\vecb h}}
\newcommand{\Lv}{{\vecb L}}
\newcommand{\Mv}{{\vecb M}}
\newcommand{\Nv}{{\vecb N}}
\newcommand{\chiv}{{\vecb\chi}}
\newcommand{\phiv}{{\vecb\phi}}
\newcommand{\etav}{{\vecb\eta}}
\newcommand{\zetav}{{\vecb\zeta}}

\newcommand{\gv}{{\vecb g}}
\newcommand{\av}{{\vecb a}}

\newcommand{\uv}{{\vecb u}}
\newcommand{\ov}{{\vecb 0}}

\begin{document}

\title{\bf Initial conditions for models of dynamical systems}

\author{Stephen~M.~Cox\thanks{Department of Theoretical Mechanics,
         University of Nottingham,
         University Park,
         Nottingham NG7 2RD,
         United Kingdom.
         E-mail: {\tt etzsmc@unicorn.nottingham.ac.uk}}
\and
         A.~J.~Roberts\thanks{Department of Mathematics \& Computing,
         University of Southern Queensland,
         Toowoomba 4350,
         Australia.
         E-mail: {\tt aroberts@usq.edu.au}}
}
\date{Sept.~14, 1994.
      To appear in {\em Physica D}.}
\maketitle



\begin{abstract}
The long-time behaviour of many dynamical systems may be effectively
predicted by a low-dimensional model that describes the
evolution of a reduced set of variables.  We consider the question of how
to equip such a low-dimensional model with appropriate initial conditions,
so that it faithfully reproduces the long-term behaviour of the original
high-dimensional dynamical system.  Our method involves putting the
dynamical system into normal form, which not only
generates the low-dimensional model, but also provides the correct
initial conditions for the model.  We illustrate the method with several
examples.
\end{abstract}

\tableofcontents

\section{Introduction}

The evolution of many physical systems may be described by ordinary
differential equations (ODEs) for the ``normal modes'' of the system. It is
often the case that most of the modes are strongly damped (these are
``stable modes'') while the others (``critical modes'') are undamped, or
nearly so. In this case, the solution of an initial-value problem rapidly
approaches a low-dimensional centre manifold (denoted by $\cal M$), which
may be parameterised by the amplitudes of the critical modes. The
subsequent evolution of the system on $\cal M$ is slower than during
the initial rapid approach to $\cal M$.

The centre manifold is an important theoretical and applicable tool for
several reasons. Firstly, we are often concerned with the stability of a
fixed point to small disturbances. The stability and bifurcations of
the fixed point may be completely determined by analysing just the
low-dimensional evolution restricted to $\cal M$ \cite{Car81,CM83}. Rational
approximations to $\cal M$ may be computed, that is, there exist {\em
constructive} schemes for approximately eliminating the stable variables
\cite{CR91,MR90,MR94,MP86,Miel88,Rob88,Rob92,Rob93}. Such schemes
involve approximating $\cal M$ by (the first few terms in) a power series
in the amplitudes of the critical modes. Secondly, if the fixed point is
stable then to any solution $P(t)$ of the original problem there
corresponds a solution $Q(t)$ on $\cal M$ which $P(t)$ approaches
exponentially quickly. The long-time behaviour of the full problem is
therefore determined by the low-dimensional (and hence tractable)
centre-manifold model.

The subject of this paper is the relationship between the full solution
$P(t)$ and its approximation $Q(t)$ on $\cM$. Specifically, we consider how
to derive the initial value $Q(0)$ from $P(0)$.
For although the theory of centre manifolds guarantees
the existence of $Q(t)$, it gives no general constructive means of
determining $Q(0)$ (a method for small $|P(0)-Q(0)|$ was given by Roberts
\cite{Rob89b}). Once $Q(0)$ is known, the solution $Q(t)$ at later times
is computed by integration of a low-dimensional system of ODEs.

Computing $Q(0)$ by a perturbation expansion in a small parameter related
to the time-scale of the approach to the centre manifold has been discussed
by van Kampen \cite[see also the references therein]{Kam85}. However, his
method requires the solution of a succession of ordinary differential
initial-value problems, which is often impractical. In Sections~2 and~3 we
develop a simpler procedure for calculating $Q(0)$, by algebraic
manipulations alone: no ODEs need be solved, as we illustrate in a simple
example in Section~4, where we compare our method with van~Kampen's. The heart
of our method is putting the evolution equations into normal
form \cite{Arn83,CS86}.
A well-known by-product of the reduction to normal form is the centre
manifold \cite{ETBCI}.  The novelty of our work lies in the observation
that appropriate initial conditions for the centre-manifold model
follow naturally from the normal-form reduction.

We further illustrate our ideas in Section~5 with Taylor's model of shear
dispersion in a channel, where both the original system and the model
are {\em partial} differential equations. Nevertheless, the model, which
involves just one spatial dimension, is considerably simpler than the
original problem, which involves two space dimensions. However, at present
there is little theory to support an infinite-dimensional centre manifold,
so the treatment is necessarily formal. As a further extension of our
approach, we discuss in an appendix the use of normal form transformations
in constructing initial conditions for dynamical models that involve both
unstable and stable modes.

\section{A system with centre and stable modes}
\label{seccm}

We consider in this section a nonlinear system of the form
\begin{eqnarray}
\dot{\xv}=A\xv+\Mv(\xv,\yv)&&\qquad \xv\in \reals^m
\label{xdot}\\
\dot{\yv}=B\yv+\Nv(\xv,\yv)&&\qquad \yv\in \reals^n,
\label{ydot}
\end{eqnarray}
where the eigenvalues $\alpha_k$ of the $m\times m$ matrix $A$ satisfy
$\ReP(\alpha_k)=0$, and the eigenvalues $\beta_k$ of the $n\times n$ matrix
$B$ satisfy $\ReP(\beta_k)<-\beta<0$. The functions $\Mv$ and $\Nv$
are strictly nonlinear smooth functions, of $\Ord{|(\xv,\yv)|^2}$.

Under these conditions centre manifold theory may be applied \cite{Car81}
to deduce the following results.
\begin{enumerate}

\item[CM1] There exists an $m$-dimensional centre manifold $\cal M$
of the form $\yv=\hv(\xv)$, with $\hv(\xv)=\Ord{|\xv|^2}$. On $\cM$
the $m$-dimensional dynamics of $\xv$ are described by~(\ref{xdot})
restricted to $\cal M$, namely
\begin{equation}
\dot{\xv}=A\xv+\Mv(\xv,\hv(\xv)).
\label{xMdot}
\end{equation}

\item[CM2] Corresponding to each solution $P(t)\equiv(\xv(t),\yv(t))$
of~(\ref{xdot}--\ref{ydot}) there is a solution $\xv=\uv(t)$ of~(\ref{xMdot}),
such that
\begin{equation}
|P(t)-Q(t)|= \Ord{e^{-\beta t}}\qquad\mbox{as $t\rightarrow\infty$,}
\label{P-Q}
\end{equation}
where $Q(t)\equiv\left(\uv(t),\hv(\uv(t))\right)$.

A sufficient condition
for this property to hold is that the origin should be stable. This condition
is not, however, necessary, and we assume that CM2 holds regardless of the
stability of the origin.

\item[CM3] If an approximation $\phiv(\xv)$ to $\hv(\xv)$ satisfies
\begin{equation}
	B\phiv+\Nv(\xv,\phiv) -\pder{\phiv}{\xv}
	\left[A\xv+\Mv(\xv,\phiv)\right]=\Ord{|\xv|^p}
	\qquad\mbox{as $|\xv|\rightarrow 0$}
	\label{happ}
\end{equation}
then $\hv=\phiv(\xv)+\Ord{|\xv|^p}$.

\end{enumerate}
Solutions of (\ref{xdot}--\ref{ydot}) therefore are exponentially attracted to
$\cM$ (at least for sufficiently small initial values of $\xv$ and
$\yv$), and at large times the essential dynamical behaviour
of~(\ref{xdot}--\ref{ydot}) is captured by~(\ref{xMdot}). This is an
important result for practical purposes because~(\ref{xMdot}) is of low
dimension, $m$, compared with~(\ref{xdot}--\ref{ydot}), which is of
dimension $m+n$.  Indeed, centre manifold theory also applies to problems with
infinitely many stable modes \cite{Car81}, for example, the
Kuramoto-Sivashinsky equation \cite{AGH89}.
Therefore, when interested only in the long-term behaviour
of~(\ref{xdot}--\ref{ydot}) we may as well compute solutions $Q(t)$
of~(\ref{xMdot}) rather than solutions $P(t)$ of~(\ref{xdot}--\ref{ydot}).
Furthermore, an important
property of~(\ref{xdot}--\ref{ydot}) is that if $|\beta_j|\gg1$ for some of
the eigenvalues $\beta_j$ then the system is stiff, which makes reliable
numerical computations expensive. The model~(\ref{xMdot}) is free of such
undesirable stiffness (although it may have the different problem of rapid
oscillations), which facilitates numerical computations, particularly over
long times.

In general $\hv(\xv)$ cannot be computed exactly. But by a straightforward
algorithm involving algebraic manipulations based on~(\ref{happ}), a
power series in $\xv$ may be developed for $\hv(\xv)$ \cite{Car81,Rob85}.
In the following subsections we describe a similarly straightforward
algorithm, based on a normal form transformation, for determining the
correct initial value $Q(0)$ from the initial value $P(0)$, so that the
exponential approach~(\ref{P-Q}) is achieved. Only with a correct initial
condition can long-term quantitative predictions be made by the
low-dimensional model~(\ref{xMdot}).

\subsection{Normal form and centre manifold}

Our method for constructing appropriate initial conditions relies on the
structure of the normal form transformation. As shown by Elphick et al.\
\cite{ETBCI}, for example, it is always possible to find a nonlinear change
of coordinates, to variables $\chiv\in\reals^m$ and $\etav\in\reals^n$, of
the form
\begin{eqnarray}
	\xv & = & \chiv+F(\chiv,\etav)\etav
	\label{xch} \\
	\yv & = & \etav+\gv(\chiv,\etav)
	\label{ych}
\end{eqnarray}
such that the dynamical evolution of~(\ref{xdot}--\ref{ydot})
may be written in normal form
\begin{eqnarray}
\dot{\chiv}&=&A\chiv+\av(\chiv)
\label{chidot}\\
\dot{\etav}&=&B\etav+\tilde B(\chiv,\etav)\etav.
\label{etadot}
\end{eqnarray}
In these equations: $F$ is an $m\times m$ matrix, $\tilde B$ is an
$n\times n$ matrix, and both are at least linear functions of their arguments.
The functions $\gv\in\reals^n$ and $\av\in\reals^m$ are strictly nonlinear.
The nonlinear terms that remain in the normal form equations
(\ref{chidot}--\ref{etadot}) arise from resonances between eigenvalues of
the linear operators \cite{Arn83}.

Note from~(\ref{etadot}) that the centre manifold $\etav=\ov$ is invariant
under the evolution of the system.  For small $\chiv$ and
$\etav$ the linear term $B\etav$ dominates the nonlinear terms
in~(\ref{etadot}), and so $\etav\rightarrow\ov$ exponentially quickly. In
terms of the original variables, $\cal M$ is given by
\[
\xv=\chiv,\qquad\yv=\gv(\chiv,\ov),
\]
or $\yv=\hv(\xv)$, where $\hv(\xv)=\gv(\xv,\ov)$. Equations~(\ref{xMdot})
and~(\ref{chidot}) describe the same long-term dynamics.

\subsection{Initial conditions}
\label{secic}

We now show that the normal form transformation gives as a by-product an
appropriate initial condition $Q(0)$, so that we can make long-term
predictions with~(\ref{xMdot}).

First we observe that if the initial value $\chiv_0$ is fixed, all solutions of
the transformed equations~(\ref{chidot}--\ref{etadot}) from initial
conditions $(\chiv_0,\etav_0)$ have the same long-term dynamics, regardless
of the value of $\etav_0$ (provided it is small enough to guarantee
approach to $\cal M$). This is because the evolution
equation~(\ref{chidot}) for $\chiv$ is independent of $\etav$. Therefore
$(\chiv_0,\ov)$ is {\em the} initial condition on $\cal M$ for a solution
that soon evolves identically to the solution from the initial
condition $(\chiv_0,\etav_0)$ off $\cal M$.

Now it is clear how to derive an initial condition $Q(0)$ for the model
(\ref{xMdot}) from the initial condition $(\xv(0),\yv(0))=(\xv_0,\yv_0)$.
\begin{enumerate}
\item  Use the coordinate transformation~(\ref{xch}--\ref{ych}) to determine
the corresponding initial condition $(\chiv_0,\etav_0)$ for the normal
form equations.
\item  As noted above, the corresponding initial condition on $\cal M$ is
$(\chiv(0),\etav(0))=(\chiv_0,\ov)$.
\item By the coordinate transformation~(\ref{xch}), the initial condition
for the model~(\ref{xMdot}) is then $\xv(0)=\chiv_0$; that is,
$Q(0)=(\chiv_0,\gv(\chiv_0,\ov))$.
\end{enumerate}

Since the change of variables is determined in powers of the new variables
$\chiv$ and $\etav$, the evolution equations of the normal form,
(\ref{chidot}) and~(\ref{etadot}), are equivalent
to~(\ref{xdot}--\ref{ydot}) only up to terms smaller than any power of
$|(\chiv,\etav)|$ as $|(\chiv,\etav)|\rightarrow0$. The differences will
become significant if our initial point is close to the boundary between
the basins of attraction of different attractors on $\cal M$ (particularly
this will be a problem if the attractors have fractal basin boundaries) or
if the solution on $\cal M$ has a positive Lyapunov exponent.
In these cases our method may, like almost any other approximation,
fail reliably to predict the long-term behaviour of the system.
An analysis of this limitation is beyond the scope of this article.

Provided the initial value of $\etav$ is not too large, all initial
conditions for~(\ref{chidot}--\ref{etadot}) in a given plane
of constant $\chiv$ evolve in essentially the same way at large
times.  In the original $(\xv,\yv)$-space, the planes
$\Pi(\chiv_0) =\left\{(\chiv,\etav) \mid\chiv=\chiv_0\right\}$
appear as curved manifolds $\Sigma$:
\[\Sigma(\chiv_0)=\left\{(\xv,\yv)\mid\xv=\chiv_0+F(\chiv_0,\etav)\etav,\
\yv=\etav+\gv(\chiv_0,\etav)\right\},
\]
where $\chiv_0$ is fixed on each manifold, and $\etav$ parameterises it.

The manifolds $\Sigma(\chiv_0)$, which Roberts~\cite{Rob89b} has termed
``isochronic manifolds'', are a generalisation of the concept of
``isochrons'' introduced by Winfree \cite{Win74} in the context of
resetting biological clocks.  Winfree considered a nonlinear oscillator
with a stable limit cycle $\Gamma$ as a model for a biological clock.
Each point on $\Gamma$ can be assigned a phase $\phi$ which increases
uniformly in time, $\dot\phi=\omega$.  If a solution on the limit cycle
is perturbed away from $\Gamma$ then it will relax back to $\Gamma$---the
biological clock will reset itself---but with a slightly altered phase.
Winfree proposed the concept of isochrons to describe how different
disturbances induce different changes to the phase of the solution.  To
each point $Q$ on $\Gamma$, with phase $\phi$ say, he associated a
surface $\Sigma(\phi)$, transverse to $\Gamma$, passing through $Q$.
This isochron $\Sigma(\phi)$ consisted of all points $P$
which, after their transient approach to $\Gamma$, have the same phase as
$Q$ has at the same time.  Later, Guckenheimer \cite{Guc75} proved the
existence of isochrons for a nonlinear oscillator, using the invariant
manifold theorem.

Our method for computing appropriate initial conditions for a
centre-manifold model, is founded on the fact that projection
along the planes $\Pi$ is simpler than nonlinear
projection along the curved surfaces $\Sigma$.

\subsection{An example}

Consider the system
\begin{eqnarray}
\dot{X}&=&X(1-X-\beta Y)\label{ex1.1}\\
\dot{Y}&=&Y(1-Y-\alpha X),\label{ex1.2}
\end{eqnarray}
which models a range of phenomena  from population dynamics to
competing modes near a multiple bifurcation point.
To demonstrate  our method we examine the degenerate case of
$\beta=1$, and make the linear change of variables
\[
X=x,\quad\quad Y=1+y-\alpha x
\]
in order to bring (\ref{ex1.1}--\ref{ex1.2}) into the form
(\ref{xdot}--\ref{ydot}), with
\begin{eqnarray}
\dot{x}&=&-(1-\alpha)x^2-xy\label{ex1.3}\\
\dot{y}&=&-y-\alpha(1-\alpha)x^2-y^2.\label{ex1.4}
\end{eqnarray}
We now apply the first nonlinear stage of a normal form transformation to
remove unnecessary quadratic terms from (\ref{ex1.3}--\ref{ex1.4}), by
setting
\begin{eqnarray}
x&=&\chi+a\chi^2+b\chi\eta+c\eta^2\label{ex1.5}\\
y&=&\eta+d\chi^2+e\chi\eta+f\eta^2\label{ex1.6},
\end{eqnarray}
with constants $a$--$f$ to be chosen.
Under this change of variable (\ref{ex1.3}--\ref{ex1.4}) become
\begin{eqnarray*}
\dot{\chi}-b\chi\eta-2c\eta^2&=&-(1-\alpha)\chi^2-\chi\eta
+\Ord{|(\chi,\eta)|^3}\\
\dot{\eta}-e\chi\eta-2f\eta^2&=&-\eta-\left[d+\alpha(1-\alpha)\right]\chi^2
				-e\chi\eta-\left[f+1\right]\eta^2
+\Ord{|(\chi,\eta)|^3}.
\end{eqnarray*}
To simplify the equations that govern $\dot\chi$ and $\dot\eta$ we choose
$b=1$, $c=0$, $d=-\alpha(1-\alpha)$, $f=1$. The coefficients $a$ and $e$
remain at our disposal and we choose to set them to zero.
The evolution equations for $\chi$ and $\eta$ are then, to quadratic
order,
\begin{eqnarray}
\dot{\chi}&=&-(1-\alpha)\chi^2\label{ex1.7}\\
\dot{\eta}&=&-\eta.\label{ex1.8}
\end{eqnarray}

Suppose now that we are given the initial condition
$P(0)=(x_0,y_0)$ for (\ref{ex1.3}--\ref{ex1.4}) near $(x_0,y_0)=(0,0)$.
Then by (\ref{ex1.5}--\ref{ex1.6}) the corresponding values for $\chi$
and $\eta$ are
\begin{eqnarray*}
\chi_0&=&x_0-x_0y_0+\Ord{|(x_0,y_0)|^3}\\
\eta_0&=&y_0+\alpha(1-\alpha)x_0^2-y_0^2+\Ord{|(x_0,y_0)|^3}.
\end{eqnarray*}
According to (\ref{ex1.8}) $\eta\rightarrow0$ exponentially as
$t\rightarrow\infty$, and $\chi$ evolves independently of $\eta$, so the
long-term evolution in (\ref{ex1.7}--\ref{ex1.8}) is the same as
from the initial condition $(\chi_0,0)$. This corresponds to the
initial condition $Q(0)=(x_0^*,y_0^*)$ for (\ref{ex1.3}--\ref{ex1.4}),
where from (\ref{ex1.5}--\ref{ex1.6})
\begin{eqnarray*}
x_0^*&=&\chi_0=x_0-x_0y_0\\
y_0^*&=&-\alpha(1-\alpha)x_0^2.
\end{eqnarray*}

\section{Slow-manifold models}

Suppose now that the rapidly decaying modes of~(\ref{xdot}--\ref{ydot}) have
been eliminated, and that the dynamics have been reduced to~(\ref{xMdot})
on the centre manifold. Some eigenvalues of $A$
represent slowly evolving modes, others fast oscillations. In this section
we discuss models in which fast oscillations are eliminated, as the rapid
transients were eliminated in the previous section. This is the essence of
many physical approximations, and is a useful tool for numerical
integrations. This procedure finds application in fluid mechanics, where
high-frequency sound waves are ignored under the incompressible
approximation \cite[\S9]{Kam85}; in beam theory, where fast ringing
modes are ignored \cite{Rob93}; in meteorology, where weather data must be
``initialised'' before numerical forecasts can be computed, in order to
remove spurious, relatively high-frequency gravity wave oscillations
\cite{CR94,Lei80,Lor86}.

Suppose the linear spectrum of $A$ consists of a zero eigenvalue repeated $m$
times, together with $n/2$ complex-conjugate pairs of purely imaginary
eigenvalues $\pm i\omega_k$. Then with some renaming of variables we may
split~(\ref{xMdot}) into slow and fast components,
\begin{equation}
\dot{\xv}=A\xv+\Mv(\xv,\yv),\qquad
\dot{\yv}=B\yv+\Nv(\xv,\yv),
\label{subcsys}
\end{equation}
where $\xv\in\reals^m$, $\yv\in\reals^n$, $A$ has $m$ eigenvalues
that are precisely zero, and $B$ has purely imaginary
eigenvalues, $(\beta_1,\ldots,\beta_n)$ with $\beta_{2k}=-\beta_{2k-1}$.
Sijbrand \cite{Sij85} shows that, since there is no resonant
forcing by $\xv$ of the fast oscillations in $\yv$, this system has a
sub-centre manifold ${\cal M}_S$ of the form $\yv=\hv(\xv)$: a ``slow
manifold'' that may be derived for this system by the same formal means
as the centre manifold described earlier. However, the slow manifold does
not in general attract neighbouring solutions, but instead acts as a slowly
evolving ``centre'' for their rapid oscillations.

For a slow-manifold model the normal form for $\dot{\chiv}$ cannot in general
be made independent of $\etav$---it takes the form
\begin{equation}
\dot{\chiv}=A\chiv+\av(\chiv,\etav),
\label{chidots}
\end{equation}
with $\av(\chiv,\etav)=O(|\etav|^2)$ as $|\etav|\rightarrow0$. Quadratic
terms in $\etav$ arise from resonant forcing of the slow modes by nonlinear
combinations of two conjugate modes $\pm i\omega_k$. The evolution equation
for the fast variables $\etav$ may still be written in the
form~(\ref{etadot}).
The subcentre manifold devoid of the fast oscillations characterised by the
imaginary eigenvalues of $B$ is given by $\etav=\ov$; the evolution of the
slow variables $\chiv$ is given by~(\ref{chidots}) restricted to $\etav=\ov$.

\subsection{Initial conditions}

Integrating~(\ref{subcsys}) numerically can be costly if the rapid
oscillations (potentially of large amplitude) are present.
Unfortunately,  since (\ref{chidots}) cannot in general be made
independent of $\etav$,  solutions on and off the slow manifold (where
$\etav=\ov$ and $\etav\neq\ov$, respectively) evolve in essentially
different ways.
It is therefore impossible in general to project an initial condition
$(\xv_0,\yv_0)$ to $\cM_S$ so as to maintain the same
essential dynamical behaviour for all time.
However, it {\em is} possible to match the behaviours for a time of
$o(|\etav|^{-2})$ in the presence of fast oscillations of small amplitude
$|\etav|$.  This approximation, where only those terms up to linear order
in $\etav$ are kept, is described by Roberts \cite{Rob89b}.

A more favourable situation occurs when the system~(\ref{subcsys}) is
symmetric.  If $\Mv(-\xv,-\yv)=-\Mv(\xv,\yv)$, and similarly for $\Nv$,
then a fast mode cannot resonate with its conjugate alone to force the
slow modes.  Instead a combination of at least three modes of at least two
different frequencies is needed to force the slow
modes.  Hence in this case $\av(\chiv,\etav)=\Ord{|\etav|^q}$ for some
$q\geq 3$ and the effects of any fast oscillations on the long-term
evolution are much weaker.

\subsection{A simple example}

Consider the system
\[
\dot{x}=y_2^2,\qquad\dot{y}_1=-y_2,\qquad\dot{y}_2=y_1,
\]
which is of the form~(\ref{subcsys}).  The $\yv$-variables execute
fast oscillations with period $2\pi$, while the variable $x$ is
constant according to the linearised equations, but is forced
quadratically by the rapid oscillations. Here the normal form transformation
can be accomplished exactly.  We set
$\chi=x+y_1y_2/2$, $\eta_1=y_1$, and $\eta_2=y_2$, so that the normal
form equations become
\[
\dot\chi=\left(\eta_1^2+\eta_2^2\right)/2,\qquad
\dot\eta_1=-\eta_2,\qquad\dot\eta_2=\eta_1,
\]
or
\begin{equation}
	\dot\chi=r^2/2,\qquad\dot r=0,\qquad\dot\theta=1,
	\label{egslow}
\end{equation}
where we have written $\eta_1=r\cos\theta$ and $\eta_2=r\sin\theta$. Thus
$r$ is the amplitude of the oscillations, and $\theta$ is
their phase. From~(\ref{egslow}), the invariant slow
manifold, ${\cal M}_S$, is given simply by $\etav=\ov$ ({\em i.e.\ }$r=0$);
the evolution
of a solution $Q(t)$ on ${\cal M}_S$ is trivial: $\dot\chi=0$. So solutions
$Q(t)$ have $\chi$ constant and $\eta_1=\eta_2=0$. However, for solutions
$P$ off ${\cal M}_S$, $r$ is a non-zero constant, and so $\chi$ drifts at a
(non-zero) constant rate: $\dot\chi=r^2/2$. It is not possible to match
solutions $P(t)$ and $Q(t)$ for large times by any choice of initial
condition $Q(0)$.

\section{A comparison of our method with others}

In order to show how the method we have described for determining
$P(0)-Q(0)$ compares with other methods, we describe now
three alternative approaches for a simple problem.
(The difference $P(0)-Q(0)$ is sometimes termed ``initial slip''
\cite{Gra63} by analogy with the boundary slip of an inviscid fluid.)
These are: in the first instance to  solve it exactly; secondly
to apply our normal-form method described above, and finally
to use explicit perturbation expansions for the {\em
solution} (rather than for the governing {\em equations}, as is the case
for the normal form method) to derive the same results.
In general, for nonlinear problems, the first approach is not available to
us, and the algebraic details of the third quickly become overwhelming.
A further method for computing the ``initial slip'', which we shall
not discuss here, has been derived by Geigenm\"uller et al.~\cite{GTF83},
who apply a systematic perturbation  procedure to linear systems of
a certain form.

The example involves a simple chemical reaction \cite[\S2]{Kam85}
which in scaled variables that represent two concentrations is written
\begin{eqnarray}
\dot{x}&=&-\epsilon(x-y)\label{vK1}\\
\dot{y}&=&kx-y          \label{vK2}.
\end{eqnarray}
In the first equation $x$ evolves slowly, because we consider $\epsilon$
to be small. These equations are linear and so the system may be solved
exactly, in particular the behaviour of the system at large times is
determined exactly.

\subsection{Exact solution}

Equations~(\ref{vK1}--\ref{vK2}) may be rewritten as
\begin{equation}
\frac{d}{dt}\left[\begin{array}{c}x\\y\end{array}\right]=
M\left[\begin{array}{c}x\\y\end{array}\right],\quad\mbox{where }
M=\left[\begin{array}{cc}-\epsilon&\epsilon\\k&-1\end{array}\right].
\label{vK3}
\end{equation}
The eigenvalues $\lambda_0$ and $\lambda_s$ of the matrix $M$ are
\[
\lambda_0,\lambda_s=\frac{1}{2}\left(-(1+\epsilon)\pm
\sqrt{(1-\epsilon)^2+4\epsilon k}\right),
\]
As $\epsilon\rightarrow0$, the eigenvalues take the form
$\lambda_0\sim\epsilon(k-1)$, $\lambda_s\sim -1$, corresponding to slow
evolution and rapid decay, respectively.

The solution of~(\ref{vK3}) is
\[
x(t)=ae^{\lambda_0t}+be^{\lambda_st},\quad
y(t)=\frac{ka}{1+\lambda_0}e^{\lambda_0t}+
  \frac{kb}{1+\lambda_s}e^{\lambda_st},
\]
where the constants $a$ and $b$ are found from the initial values:
\[
\left[\begin{array}{c}a\\b\end{array}\right]=
\frac{(1+\lambda_0)(1+\lambda_s)}{k(\lambda_0-\lambda_s)}
\left[\begin{array}{c}-y_0+kx_0/(1+\lambda_s)\\
                       y_0-kx_0/(1+\lambda_s)\end{array}\right].
\]
Since $1\gg\epsilon(k-1)$, the $\lambda_s$-mode decays to zero
much more rapidly than the $\lambda_0$-mode evolves, so that after some
time the solution is given, {\em to within exponentially small terms,} by
\begin{equation}
x(t)\sim ae^{\lambda_0t},\quad
y(t)\sim \frac{ka}{1+\lambda_0}e^{\lambda_0t}=\frac{k}{1+\lambda_0}x
\label{xy}
\end{equation}
This solution is equivalent to the solution of the
one-dimensional model equation
\begin{equation}
\dot{X}=\lambda_0X\label{X}
\end{equation}
with initial condition
\begin{equation}
X(0)=\frac{(1+\lambda_0)(1+\lambda_s)}{k(\lambda_0-\lambda_s)}
\left(-y_0+\frac{kx_0}{1+\lambda_s}\right),
\label{ic1}
\end{equation}
where $x=X$ and $y=kX/(1+\lambda_0)$.

\subsection{Normal form calculation}

The normal form for~(\ref{vK1}--\ref{vK2}) is obtained by making the
substitution
\begin{equation}
x=\chi+f\eta,\quad y=\eta+g\chi,
\label{nfex}
\end{equation}
where $f(\epsilon)$ and $g(\epsilon)$ are constants. Wycoff \& Balasz
\cite{WB87} have considered linear systems of a form that includes
(\ref{vK1}--\ref{vK2}), and have derived a substitution of the form
(\ref{nfex}) by considering a multiple-time-scale perturbation expansion.
Our results for this problem agree with theirs for the more general case.
Equations~(\ref{vK1}--\ref{vK2}) fail to satisfy the
conditions of~(\ref{xdot}--\ref{ydot}) in two ways: firstly, there is no
mode with precisely zero growth rate; secondly, the system has not been
decomposed into critical and stable subspaces.
The first point is dealt with by using a standard
trick of bifurcation theory \cite{Car81}, namely considering the parameter
$\epsilon$ as an extra dynamical variable with the trivial evolution
equation
\begin{equation}
	\dot{\epsilon}=0.
	\label{epsdot}
\end{equation}
Then the term $-\epsilon(x-y)$ is {\em nonlinear} in the extended
system~(\ref{vK1},\ref{vK2},\ref{epsdot}),
and so the linearised evolution of $x$ is just $\dot{x}=0$.  The second
point is unimportant in practice, and we diagonalise the system as we go.

Substituting~(\ref{nfex}) into the governing equations~(\ref{vK1}--\ref{vK2})
we find
\begin{eqnarray*}
\dot{\chi}+f\dot{\eta}&=&
          -\epsilon\chi-\epsilon f\eta+\epsilon\eta+\epsilon g\chi\\
\dot{\eta}+g\dot{\chi}&=&
          -\eta-g\chi+k\chi+kf\eta.
\end{eqnarray*}
Separating out terms proportional to $\chi$ and $\eta$ we find that the
slow and fast variables evolve according to
\begin{eqnarray}
\dot{\chi}&=&-\epsilon(1-g)\chi
\label{exchi1} \\
\quad\dot{\eta}&=&-(1-kf)\eta,
\label{exchi2}
\end{eqnarray}
where $f$ and $g$ must satisfy quadratic equations whose roots are
\begin{eqnarray*}
f&=&\frac{1}{2k}\left(1-\epsilon-\sqrt{(1-\epsilon)^2+4\epsilon
k}\right) \sim-\epsilon\quad\quad \mbox{as }\epsilon\rightarrow0\\
g&=&\frac{1}{2\epsilon}\left(\epsilon-1+\sqrt{(1-\epsilon)^2+4\epsilon
k}\right) \sim k\quad\quad \mbox{as }\epsilon\rightarrow0.
\end{eqnarray*}
The fact that $g$ is not small as $\epsilon\rightarrow0$
reflects the fact that the original system is not diagonalised.

The centre manifold may now be read off from the normal form~(\ref{nfex})
by setting $\eta=0$. That is,
\begin{equation}
x=\chi,\quad y=g\chi.
\label{xynf}
\end{equation}
Since $g(1+\lambda_0)=k$, this expression for the centre manifold agrees
with~(\ref{xy}) derived from the exact solution (as it should).
Since $-\epsilon(1-g)=\lambda_0$, the evolution of the slow variable
$\chi$ in~(\ref{exchi1}) is precisely that derived from the exact
solution, where we denoted the slow variable by $X$ in~(\ref{X}).

Now we follow the procedure described in Section~\ref{secic} to determine the
appropriate initial condition $Q(0)$  on the centre manifold which gives rise
to the same long-term dynamics as $P(0)=(x_0,y_0)$. Using~(\ref{nfex})
we find that $P(0)$ is written in terms of $\chi$ and $\eta$ as
\[
(\chi_0,\eta_0)=\frac{1}{1-fg}(x_0-fy_0,-gx_0+y_0).
\]
The expression for the corresponding $Q(0)$ on $\cM$ is $(\chi_0,0)$.
Transforming this point back to the original variables yields the expression
\begin{equation}
(x^*_0,y^*_0)=\frac{1}{1-fg}\left(x_0-fy_0,g[x_0-fy_0]\right).
\label{nfic}
\end{equation}
A little algebra shows that (\ref{nfic}) agrees with the exact
solution~(\ref{ic1}).

\subsection{Perturbation expansion of the solution}

Finally, we treat the problem with a techniquebased upon a perturbation
expansion for the solution during its rapid approach to the centre
manifold. Such a method is described by van Kampen \cite{Kam85} for
problems that may be nonlinear, and we follow his argument here.
The essential idea is that during the rapid approach to the centre
manifold, the fast variables change a great deal, while the slow
variables change little.

We begin by expanding the slow variable~$x(t)$ as a power series in~$\epsilon$,
\begin{equation}
x(t)=x^0(t)+\epsilon x^1(t)+\epsilon^2 x^2(t)+\cdots,
\label{ser}
\end{equation}
and then we substitute this expansion into~(\ref{vK1}--\ref{vK2}). Note the
important difference between this perturbation expansion of the {\em
solution} and the expansion of the {\em governing equations} for the
normal form calculation. Approximations to the fast variable~$y(t)$ are
constructed from~(\ref{vK2}), with $x$ replaced by a finite truncation
of the series~(\ref{ser}).

At $\Ord{\epsilon^0}$ in~(\ref{vK1}), $\dot{x}^0=0$, so
\begin{equation}
x^0=\mbox{constant}=x_0.
\label{x0}
\end{equation}
To leading order then, during the approach of the solution to $\cal M$,
$x$ does not change. Now we compute an approximation to $y(t)$ during
the approach to $\cal M$ using~(\ref{x0}) and~(\ref{vK2}),
\[
\dot{y}=-y+kx^0,
\]
which may be solved to give $y(t)=y_0e^{-t}+kx_0(1-e^{-t})$.
Now we usethis solution for $y(t)$ to compute $x_1$,
by considering~(\ref{vK1}) at~$\Ord{\epsilon^2}$:
\[
\dot{x}^1=-x^0+y(t),
\]
whose solution is
\[
x^1(t)=-x_0t+y_0(1-e^{-t})+ky_0(t+e^{-t}-1).
\]
Now we re-assemble the solution $x(t)$ to $\Ord{\epsilon}$ and find
\[
x(t)\sim \left\{x_0+\epsilon(y_0-kx_0)\right\}+
     \epsilon t\left\{-(1-k)x_0\right\}+
     \epsilon e^{-t}\left\{-y_0+kx_0\right\}.
\]
The first two terms are the start of a Taylor expansion of the solution in
powers of $t$, while the last term is an exponentially decaying transient.
The initial value for the Taylor-expansion component,
$x_0+\epsilon(y_0-kx_0)$, is just the initial value for the solution
$Q(t)$ on $\cal M$.
This result agrees with the previous two calculations of the initial
condition $Q(0)$, as we see by expanding~(\ref{nfic}) to give
\[
x^*_0=\frac{x_0-fy_0}{1-fg}\sim\frac{x_0+\epsilon y_0}{1+\epsilon k}
= x_0+\epsilon(y_0-kx_0)+\Ord{\epsilon^2}.
\]
To continue the expansion requires the solution of a succession
of ODEs: in general, even if possible, the computations rapidly
become exhausting.

\section{An infinite-dimensional example: shear dispersion}
\label{sheardiseg}

We now apply the technique we have described for determining initial
conditions to a
generalisation of Taylor's model \cite{Tay53} for shear dispersion. This
model idealises the spreading of some contaminant released into a river.
The river is modelled as a channel, and the water is supposed to flow
parallel to the banks, with the flow being fastest closest to the centre of
the channel. The variation in water speed with distance from the bank
increases contaminant gradients, while molecular diffusion tends to smooth
gradients out. A balance between these competing physical mechanisms occurs
at large times after the release of the contaminant; eventually the
contaminant is approximately evenly spread in any cross-section of the
river. The cross-sectionally averaged concentration varies slowly in the
downstream direction. The slow evolution of the averaged concentration
obeys a simplified model equation which was derived through the techniques
of centre manifold theory by Mercer \& Roberts \cite{MR90,MR94}.

The following analysis can be made rigorous in Fourier space \cite{MR90},
but this rigour is generally unavailable in other problems where the
centre manifold itself is of infinite dimension.  In the absence of
rigorous theory, the best we can do is to apply our techniques formally.

\subsection{The model and its linear behaviour}

Consider the dispersion of a contaminant, with concentration $c(x,y,t)$, in a
channel $0<y<1$, $-\infty<x<\infty$. If longitudinal spatial variations
(that is, variations in $x$) occur on a large scale than $c$ satisfies the
partial
differential equation (PDE)
\begin{equation}
\pder{c}{t}=-u(y)\pder{c}{x}+\pderder{c}{y},
\label{disp}
\end{equation}
subject to the boundary conditions that no contaminant escapes through
the walls of the channel,
\begin{equation}
\pder{c}{y}=0\quad\mbox{at}\quad y=0,1.
\label{bcy}
\end{equation}
The variables have been made dimensionless, and $u(y)$ is the
velocity in the $x$-direction.

The PDE~(\ref{disp}) has two space dimensions; the ``centre manifold''
we derive by performing a normal form transformation involves just one space
dimension.  We concentrate on the {\em structure} of the normal form
transformation; the details have been discussed elsewhere \cite{MR90,MR94}.

The problem~(\ref{disp}--\ref{bcy}) is linear in the concentration.
The concentration evolves so that after some time the spatial variations in
$x$ are small, and so we treat all $x$-derivatives as
``nonlinear'' terms in the same way that a small parameter was treated earlier.
With this interpretation, the ``linearised'' dynamics are given by
\[
\pder{c}{t}=\pderder{c}{y},
\]
which physically describes cross-channel diffusion.
The eigenmodes $e^{\sigma_n t}c_n(y)$ are of the form
\[
c_n=\cos n\pi y\quad (n=0,1,\ldots),
\]
with corresponding eigenvalues $\sigma_n=-n^2\pi^2$.
According to this ``linear'' picture, all modes decay except $c_0$. The
system is therefore analogous to~(\ref{xdot}--\ref{ydot}), with one
critical mode but infinitely many damped modes. Note that the critical
space is spanned by the single mode, $c_0$, and that all other eigenmodes
have zero $y$-average. For a given concentration field $c$ we can determine
the component in the critical space by taking a $y$-average to give $\avc$;
then the component in the stable space is $\cpr\equiv c-\avc$.

\subsection{Normal form}

Here we derive a ``normal form'' for~(\ref{disp}). The normal form
transformation consists of choosing new variables $\chi$ and $\eta$ in
terms of which the governing PDE~(\ref{disp}) is simplified. The general
form of this transformation is
\[
\avc=\chi+g_1(\chi,\eta),\quad \cpr=\eta+g_2(\chi,\eta).
\]

Since the governing equation is linear in $c$ then so are $g_1$ and $g_2$.
They are ``nonlinear'' in the sense that they vanish when $x$-derivatives
are ignored. Using experience from the previous sections, we choose $\avc$
to reduce to $\chi$ when $\eta=0$, and $\cpr$ to reduce to $\eta$ when
$\chi=0$. Furthermore, we suppose
that $\chi$ has no component in the stable space ($\overline{\chi}=\chi$) and
that $\eta$ has no component in the centre space ($\overline{\eta}=0$).
Such considerations imply that the ``nonlinear'' terms in the normal form
transformation must be of the form $g_1(\chi,\eta)=\theta\eta$,
$g_2(\chi,\eta)=h\chi$, where $h$ and $\theta$ are linear operators such that
$\overline{\theta\eta}=\theta\eta$ and $\overline{h\chi}=0$.
For definiteness we define $\theta$ so that $\theta\overline{f}=0$
for all $f$.  We have therefore decomposed $c(x,y,t)$ as follows:
\begin{equation}
c=\chi+h\chi+\eta+\theta\eta.
\label{cdec}
\end{equation}

Because the original problem is linear in $c$ we may deal with
the $\chi$- and the $\eta$-components separately.
After substituting~(\ref{cdec}) into~(\ref{disp}) we consider first
those terms involving $\chi$.  We find that
\begin{equation}
(1+h)\pder{\chi}{t}=-u\pder{}{x}[(1+h)\chi]+\pderder{}{y}h\chi.
\label{chiterms}
\end{equation}
This equation serves to define both the operator $\chi$ and
the derivative $\partial\eta/\partial t$.
The average of this equation over $y$ gives an expression for
$\partial\chi/\partial t$:
\begin{equation}
\pder{\chi}{t}=-\overline{u(1+h)}\pder{\chi}{x}.
\label{sdcm}
\end{equation}
This is the evolution equation for $\chi$ on the centre manifold, and
represents a simplification  of the original problem~(\ref{disp}).
Not only is~(\ref{sdcm}) of lower dimension than~(\ref{disp}), because
there is no $y$-dependence in $\chi(x,t)$, but there
are only ``nonlinear'' terms in~(\ref{sdcm})---that is, every term on the
right-hand side of~(\ref{sdcm}) is differentiated at least once with
respect to $x$.

Substituting~(\ref{sdcm}) back into~(\ref{chiterms}), we obtain an equation
that governs the operator $h$:
\begin{equation}
-(1+h)\overline{u(1+h)}\pder{}{x}=-u(1+h)\pder{}{x}+\pderder{}{y}h.
\label{heqn}
\end{equation}
In order to solve this equation, we expand $h$ in powers of
$\partial_x$,
\[
h\sim\sum_{j=1}^\infty\alpha_j(y)\partial_x^j,
\]
substitute this expression into~(\ref{heqn}), and equate powers of
$\partial_x$.
Retaining only the first term in the expansion of $h$, we find~(\ref{sdcm})
to be approximately
\[
\pder{\chi}{t}=-\overline{u}\pder{\chi}{x}-\overline{u\alpha_1}
\frac{\partial^2\chi}{\partial x^2}.
\]
The concentration $\chi$ is advected with the mean velocity,
$\overline{u}$, and diffuses with diffusion coefficient
$-\overline{u\alpha_1}$.

A similar argument for the terms involving $\eta$ yields
\[
(1+\theta)\pder{\eta}{t}=-u\pder{}{x}\left[(1+\theta)\eta\right]
+\pderder{\eta}{y}.
\]
Averaging with respect to $y$ gives
\[
\theta\pder{\eta}{t}=-\pder{}{x}\overline{u(1+\theta)\eta}.
\]
Subtracting one equation from the other gives an evolution equation for $\eta$:
\begin{equation}
\pder{\eta}{t}=\pder{}{x}\left[\overline{u\eta}-u\eta+
                               \overline{u\theta\eta}-u\theta\eta\right]+
               \pderder{\eta}{y},
\label{sdst}
\end{equation}
and so $\theta$ satisfies
\[
\theta\left\{-\pder{}{x}\left[u(1+\theta)\eta\right]+\pderder{\eta}{y}\right\}
=-\pder{}{x}\overline{u(1+\theta)\eta}.
\]
This equation may be solved by expanding $\theta$ in powers of $\partial_x$.
Note that the manifold $\eta=0$ is invariant (from~(\ref{sdst})) and
the evolution equation~(\ref{sdst}) for $\eta$  contains both
``linear'' and ``nonlinear'' terms. Since $\overline{\eta}=0$ then the
eigenvalues of the ``linear'' operator are $\lambda_1$, $\lambda_2$,\ldots,
which are all negative. Consequently, $\eta\rightarrow0$ exponentially
quickly.

If we compute the terms in the expansions for $h$ and $\theta$ we find that
$\theta\eta=-\overline{h\eta}$,
so the normal form transformation~(\ref{cdec}) puts $c$ into the form
\begin{equation}
c=\chi+h\chi+\eta-\overline{h\eta}.
\label{decomp}
\end{equation}
Note that when we apply the operator $1+h$ to~(\ref{decomp})
and average with respect to $y$ we obtain
\begin{equation}
\overline{(1+h)c}=\overline{(1+h)^2\chi}.
\label{cchi}
\end{equation}
This enables us to determine $\chi$ from a given $c$. To then determine
the corresponding value for $\eta$, we use
\begin{equation}
\eta=c-\avc-h\chi.
\label{ceta}
\end{equation}

\subsection{An appropriate initial condition}

To derive initial conditions for $\chi$ so that its long-term evolution
in~(\ref{sdcm}) agrees asymptotically with that of $c$ in~(\ref{disp})
from a given initial condition $c=c_0(x,y)$ we first
apply~(\ref{cchi}--\ref{ceta}) to obtain
the values $(\chi_0,\eta_0)$ that correspond to $c_0$. Then we note that the
solution of~(\ref{sdcm}--\ref{sdst}) from $P(0)=(\chi_0,\eta_0)$
exponentially approaches the solution from $Q(0)=(\chi_0,0)$.
Transforming $Q(0)$ back to the physical variable $c$, using~(\ref{cchi}),
we obtain a modified initial condition, $c^*_0$, for~(\ref{disp}) that lies
on the centre manifold and has the same long-time behaviour as from $c_0$.
Of course, if we want to integrate the model equation~(\ref{sdcm}) instead
of~(\ref{disp}) itself, then all we need is the value $\chi_0$.

\subsection{Comparison with systematic adiabatic elimination}

In order to emphasise the relative simplicity of our method for
determining ``initial slip'' by purely algebraic manipulations,
we now describe an alternative method (Haake and Lewenstein \cite{Haa83}):
that of systematic adiabatic elimination.
(See also a treatment of (\ref{disp}) in a different context by
Titulaer~\cite{T80}.)
The governing equation~(\ref{disp}) for the concentration $c(x,y,t)$
is written in the form
\begin{equation}
\pder{c}{t}=\left(L_0+L_1\right)c=Lc,
\label{Lc}
\end{equation}
where for the shear dispersion problem
\[
L_0=\pderder{}{y},\quad L_1=-u(y)\pder{}{x}.
\]
Formally, the solution may be written in terms of the initial concentration
distribution as $c(x,y,t)=e^{Lt}c(x,y,0)$.

An evolution equation is sought for a ``reduced'' variable
$C(x,t)=\overline{c}$.
This equation will be of the form
\[
\pder{C}{t}=\ell C,
\]
for which the formal solution, subject to the initial condition
$C(x,0)=C_\eff (x)$ is $C(x,t)=e^{\ell t}C_\eff(x)$.
Our goal is to calculate the operator $\ell$, and to determine
$C_\eff (x)$ so that the reduced description of the dynamics matches the
full solution, that is,
\[
\overline{e^{Lt}c(x,y,0)}\sim e^{\ell t}C_\eff (x)
\]
for large times. In the notation of the previous subsections
$\ell=-\overline{u(1+h)}\partial_x$ and $C_\eff (x)=\chi(x,0)$.

The computation of these quantities by systematic adiabatic
elimination proceeds as follows. First we use~(\ref{Lc}) and the
definition of $C$ to write
\begin{equation}
C(t)=\overline{c(t)}=\overline{c(0)}+\overline{\int_0^tLc(\tau )d\tau }
=\overline{c(0)}+\overline{\int_0^tLe^{L\tau }c(0)d\tau },
\label{C(t)}
\end{equation}
where we have suppressed the dependence of $c$ and $C$ on $x$ and $y$ for
notational brevity.
Now we assume that $\partial_x$, and so $L_1$, is small,
and we carry out a small-$t$, small-$L_1$ expansion for $C$. The
first step is to approximate $\exp(Lt)$ by $\exp(L_0t)$. Then we note
the decomposition
\[
e^{L_0t}c(0)=c(0)+\left[e^{L_0t}c(0)-c(0)\right].
\]
The term in brackets is just
\[
L_0^{-1} \pder{}{t}e^{L_0t}c(0),
\]
where the ``inverse'' of $L_0$ has been taken to have zero $y$-mean.
Therefore, collecting these results and substituting them into~(\ref{C(t)}),
we find
\begin{eqnarray*}
C(t)&\approx&\overline{c(0)}+\overline{\int_0^tL_1c(0)d\tau }+
\overline{\int_0^\infty L_1L_0^{-1}\pder{}{t}e^{L_0\tau }c(0)d\tau }\\
&\approx&\overline{c(0)}+t\overline{L_1c(0)}-\overline{L_1L_0^{-1}c(0)}+\cdots.
\end{eqnarray*}
Here we have extended the range of one of the integrals to infinity,
and incur exponentially small errors as a result.
This expression represents the Taylor series expansion of $C(t)$ for
small $t$, with initial condition
\begin{eqnarray*}
C_\eff (0)&=&\overline{c(0)}-\overline{L_1L_0^{-1}c(0)}.
\\
&=&\overline{c(0)}+\pder{}{x}\overline{u(y)L_0^{-1}c(0)}\\
&=&\overline{c(0)}+\pder{}{x}\overline{c(0)L_0^{-1}u(y)}.
\end{eqnarray*}
This result agrees with the previous calculation (where we use the result that
$\alpha_1=L_0^{-1}u(y)$ \cite{MR90,MR94}).

To solve for even this leading-order contribution
to the initial slip (there are also contributions from all higher powers
of $\partial_x$) we had to evaluate an integral of the
exponential of some operator. For shear dispersion, the integral can be
evaluated explicitly, although at higher orders and in
truly nonlinear problems the exact computation of such an integral will
not be possible. By contrast, the method we have proposed for computing
initial slip by purely algebraic manipulations is much more straightforward.

\section{Conclusion}

We have described a method that uses a normal form transformation to
determine appropriate initial conditions for low-dimensional models of
evolving systems. The nonlinear coordinate transformation allows one to
compute the initial slip by solving a succession of algebraic problems,
rather than the differential problems associated with other methods. The
process has been illustrated with several examples. Using a normal form
transformation is simpler than other methods, and also shows, for example,
how nonlinear resonances make it impossible, in general, to find good
initial conditions for slow sub-centre manifold models such as the
quasi-geostrophic approximation or beam theory.

Generally, the calculation of the initial slip will be carried out only
approximately (usually by resorting to the truncation of a power series),
so the solutions of the model and the full system will differ to some
order, after a period of rapid exponential approach. Since the change in
variables is made as a power series in the new variables, the approximate
normal form equations differ from their exact counterparts by, say, terms
of order $|(\chiv,\etav)|^N$ as $|(\chiv,\etav)|\rightarrow0$. The
differences will become significant if the initial point is close to the
boundary between the basins of attraction of different attractors on $\cM$,
or if the attractor on $\cM$ has a positive Lyapunov exponent. However, these
are inhospitable circumstances to which to expose our technique, and we
have shown elsewhere \cite{CR94} that for a simple
atmospheric model the procedure we have described for determining the
initial slip produces a very good agreement between $P(t)$ and $Q(t)$ over
a reasonable time-scale for forecasting, even when only the first few terms
in the power series are kept.

\appendix

\section{Other invariant manifolds}

In this appendix we generalise the statements we have so far made about
systems with a centre manifold to other types of invariant manifolds.  In
the context of forming low-dimensional dynamical models this is useful
because, for example:
some systems, when linearised, also have eigenvalues with
positive real parts \cite{AGH89};
some decaying modes may not decay fast enough to be
reasonably eliminated in a given application \cite{Rob89a,WR94}.

\subsection{Including linearly unstable modes}

First consider the case of a system where some eigenvalues of the linear
operator have positive real parts:
\begin{eqnarray}
\dot{\xv}&=&A\xv+\Mv(\xv,\yv,\zv)
\quad \xv\in \reals^m \nonumber\\
\dot{\yv}&=&B\yv+\Nv(\xv,\yv,\zv)
\quad \yv\in \reals^n \label{csueqns}\\
\dot{\zv}&=&C\zv+\Lv(\xv,\yv,\zv)
\quad \zv\in \reals^\ell,\nonumber
\end{eqnarray}
where the eigenvalues $\gamma_k$ of $C$ satisfy
$\Gamma\ge\ReP(\gamma_k)>\gamma>0$, those of $A$ satisfy
$\ReP(\alpha_k)=0$, and those of $B$ satisfy $\ReP(\beta_k)<-\beta<0$.
The normal form evolution equations are of the form
\begin{eqnarray}
\dot{\chiv}&=&A\chiv+\av(\chiv)+
                  \tilde{\av}(\chiv,\etav,\zetav)\nonumber\\
\dot{\etav}&=&B\etav+\tilde B(\chiv,\etav,\zetav)\etav\label{csunorm}\\
\dot{\zetav}&=&C\zetav+\tilde C(\chiv,\etav,\zetav)\zetav,\nonumber
\end{eqnarray}
where
$\av(\ov)=\tilde{\av}(\chiv,\ov,\zetav)=\tilde{\av}(\chiv,\etav,\ov)=\ov$.

Immediately we identify five non-trivial invariant subspaces of this system
of equations:
\begin{itemize}
	\item  the centre space with $\etav=\zetav=\ov$;

	\item   the stable space with $\chiv=\zetav=\ov$;

	\item   the unstable space with $\chiv=\etav=\ov$;

	\item   the centre-unstable space with $\etav=\ov$;

	\item   the centre-stable subspace with $\zetav=\ov$.
\end{itemize}
In terms of the original variables these five spaces respectively form five
invariant manifolds: a centre manifold $\cM$; a stable manifold $\cM_S$; an
unstable manifold $\cM_U$; a centre-unstable manifold $\cM_{CU}$; and a
centre-stable manifold, $\cM_{CS}$.

In the context of constructing low-dimensional models which capture the
long-term dynamics of the system, only one of these five invariant
manifolds is of real interest.  Solutions on the centre manifold,
the centre-stable manifold and the stable manifold are all unstable to
perturbations in the linearly unstable variables $\zetav$
near the origin.  Solutions on the unstable manifold, while stable
to perturbations in $\etav$,  may be unstable to perturbations in
the centre variables $\chiv$.  This leaves the centre-unstable manifold
$\cM_{CU}$ which is invariant and, at least near the origin, attracts
all solutions in its neighbourhood at a rate like $e^{-\beta t}$.

On the centre-unstable manifold, the dynamical behaviour is described by
\begin{equation}\label{cunorm}
\dot{\zetav}=C\zetav+\tilde C(\chiv,\ov,\zetav)\zetav\ ,
\qquad
\dot{\chiv}=A\chiv+\av(\chiv).
\end{equation}
Unlike the case without linearly unstable modes, the normal
form~(\ref{csunorm}) does not in general allow the derivation
of initial conditions for a solution $Q(t)$ on
$\cM_{CU}$ that matches a solution $P(t)$ off $\cM_{CU}$. The reason is
possible resonances between the stable and the unstable modes, which
only affect the short-term transients (as $\etav\to\ov$) but which cannot
be removed from the normal form equations~(\ref{csunorm}). This is most
easily seen in a simple example.

Consider the normal form equations
\begin{equation}\label{cutoy}
\dot\chi=\eta\zeta^2,\qquad
\dot\eta=-2\eta,\qquad
\dot\zeta=\zeta-\zeta^3.
\end{equation}
These equations have the centre-unstable manifold $\eta=0$ which attracts
all neighbouring solutions as $\eta=\eta_0\exp(-2t)\rightarrow0$.
Therefore any neighbouring solution $P(t)$ will
asymptote exponentially quickly to a solution $Q(t)$ on $\cM_{CU}$; hence
for all neighbouring initial conditions $P(0)$ there is a
corresponding initial condition $Q(0)$ on $\cM_{CU}$. By solving~(\ref{cutoy})
explicitly we find that the solution starting from
\[
\chi^*=\chi_0-\frac{\eta_0\zeta_0^2\log\zeta_0}{1-\zeta_0^2},\quad
\eta^*=0,\quad
\zeta^*=\zeta_0
\]
has the same long-term dynamics as that starting from
$(\chi_0,\eta_0,\zeta_0)$.


The presence of the logarithm in this expression is novel. It indicates
that the assumed power-series representation for the normal form
transformation is not sufficiently general for our purposes. Indeed if we
make the nonlinear change of variable
\[
 \theta=\chi-\frac{\eta\zeta^2\log\zeta}{1-\zeta^2}
\]
and leave $\eta$ and $\zeta$ unchanged then~(\ref{cutoy}) becomes
\[
  \dot\theta=0,\qquad\dot\eta=-2\eta,\qquad\dot\zeta=\zeta-\zeta^3\ .
\]
The new centre variable $\theta$ is not forced by any resonant term, and the
projection of initial conditions onto the centre-unstable manifold is
trivial.

Let $\Gamma$ be a
realistic upper bound on $\ReP(\gamma_k)$ of $C$ and let $-\beta$ be a
realistic upper bound on $\ReP(\beta_k)$ of $B$ (see text just
after~(\ref{csueqns})).
Then if $\Gamma>\beta$, that is, if the most rapidly
growing mode grows faster than the slowest decaying mode decays,
just one of the decaying modes may
directly interact with just one of the growing modes---the resonance
could be as direct as a second-order $\eta_i\zeta_j$ term.
However, if $\Gamma<\beta$, that is, if the slowest decay occurs more
rapidly than the most rapid growth, then any resonance must be of order
$\eta_i\zeta_j^{\beta/\Gamma}$ or higher. Thus the normal form projection
of initial conditions can be carried out to this order of
accuracy. Observe that as $\Gamma\to0$, when the centre-unstable manifold
becomes a centre manifold, $\beta/\Gamma\to\infty$, and the order of the
resonance moves out to infinity. This limit recovers the case of a centre
manifold where there is no problem in computing an appropriate initial
condition $Q(0)$.

\subsection{Keeping some decaying modes}

In a particular application it may be that some of the linearly stable
modes of~(\ref{csueqns}) are dynamically significant,
for example, if their decay is relatively slow.  An example
is shear dispersion in a channel or pipe, where the slowest
modes decay on a cross-stream diffusion time-scale that may
be comparable with other physical processes of interest.  We then
try to construct an invariant manifold $\cM_I$ that includes not only
the critical $\xv$-modes together with the unstable modes $\zv$,
but also the slower-decaying of the $\yv$-modes.  For shear
dispersion this was done by Watt \& Roberts \cite{WR94}.

After renaming the $\zv$ variables and some of the $\yv$ variables
of~(\ref{csueqns}) as $\xv$ variables, we consider a dynamical system
given by~(\ref{xdot}--\ref{ydot}) but where the eigenvalues $\alpha_k$ of
$A$ and those of $B$ satisfy
\begin{equation}\label{eigconst}
\ReP(\beta_{k'})<-\beta<-\gamma<\ReP(\alpha_k)<\Gamma
\quad\quad\beta,\gamma>0.
\end{equation}
Again without loss of generality we take $A$ and $B$ to be diagonal
matrices. The variables $\xv$ are termed {\em master} variables in this
context \cite{Hak78}, while the variables $\yv$ are termed {\em slave}
variables.

As before, we make a nonlinear change of variables.
Resonance may now generate a term that includes
purely master variables as a forcing in the slave equations.
Thus, even in the normal form, the slaved variables do not necessarily
evolve quickly to zero.  This means that they may evolve non-trivially at
large times; that is, on the long-term time-scale of the
master variables (``short-term'' then refers to the time-scale of the slave
variables' decay).  A resonance occurs at lowest order when $j$ master
modes combine to force one slave mode; using the bounds~(\ref{eigconst})
on the eigenvalues, the lowest order at which the resonance can occur is
\[
j_s=\beta/\gamma\ .
\]
This resonance is symptomatic of the non-uniqueness of invariant
manifolds. The non-uniqueness first appears at this order; as discussed by
Roberts \cite{Rob89a} it appears as a zero divisor in the construction of
the invariant manifold --- there is little point in computing
higher orders in the expansions because the higher order differences are
smaller than those ignored by the model.

Resonant interactions between purely
master modes are the important interactions in the long term.
However, any interaction with a slave mode results in the undesirable
feature that during the evolution onto ${\cal M}_I$ the master dynamics
depend upon the slave dynamics---an effect which we want to ignore.
The lowest order at which this resonance may appear is if one slave
mode interacts with $j-1$ of the most rapidly growing master modes to
force the most rapidly decaying master mode.  Thus the lowest order for
this resonance is
\[
j_m=1+\frac{\beta-\gamma}\Gamma.
\]
Such resonance makes it hard to prescribe correct initial conditions on the
low-dimensional manifold $\cM_I$ for any given initial condition of the
full system. However, in many examples, such as~(\ref{cutoy}), this
reflects the limitations of the standard power-series representation.

Observe that these effects show up at high-order only if the
{\em spectral gap} between the master and the slave modes,
$[-\beta,-\gamma]$, is large, both relatively (large $\beta/\gamma$)
and absolutely (large $\frac{\beta-\gamma}\Gamma$), when compared
to the natural growth rates in the dynamics.  Thus such a large spectral
gap is desirable in constructing low-dimensional dynamical models.


\begin{thebibliography}{99}
\addcontentsline{toc}{section}{References}

\bibitem{AGH89}
D.~Armbruster, J.~Guckenheimer and P.J. Holmes,
Kuramoto-Sivashinsky dynamics on the center-unstable manifold,
{\em SIAM J.\ Appl.\ Math.\ }{\bf 49} (1989) 676--691.

\bibitem{Arn83}
V.I. Arnold,
{\em Geometrical methods in the theory of ordinary differential equations\/}
(Springer, 1983).

\bibitem{Car81}
J. Carr,
{\em Applications of centre manifold theory\/}
(Springer, 1981).

\bibitem{CM83}
J. Carr and R.G. Muncaster,
The application of centre manifold theory to amplitude expansions.
I. Ordinary differential equations,
{\em J.\ Diff.\ Eqns.\ }{\bf 50} (1983) 260--279.

\bibitem{CR91}
S.M. Cox and A.J. Roberts,
Centre manifolds of forced dynamical systems,
{\em J. Austral.\ Maths.\ Soc.\ Ser.\ B} {\bf 32} (1991) 401--436.

\bibitem{CR94}
S.M. Cox and A.J. Roberts,
Initialisation and the quasi-geostrophic slow manifold,
{\em preprint} (1994).

\bibitem{CS86}
R. Cushman and J.A. Sanders,
Nilpotent normal forms and representation theory of $sl(2,R)$,
{\em Contemp.\ Math.\ }{\bf 56} (1986) 31--51.

\bibitem{ETBCI}
C.~Elphick, E.~Tirapegui, M.E.~Brachet, P.~Coullet, and G.~Iooss,
A simple global characterization for normal forms of singular vector fields,
{\em Physica D} {\bf 29} (1987) 95--127.

\bibitem{GTF83}
U.~Geigenm\"uller, U.M.~Titulaer, and B.U.~Felderhof,
Systematic elimination of fast variables in linear systems,
{\em Physica A} {\bf 119} (1983) 41--52.

\bibitem{Gra63}
H. Grad,
Asymptotic theory of the Boltzmann equation,
{\em Phys.\ Fluids} {\bf 6} (1963) 147--181.

\bibitem{Guc75}
J. Guckenheimer,
Isochrons and phaseless sets,
{\em J.\ Math.\ Biol.\ }{\bf 1} (1975) 259--273.

\bibitem{Haa83}
F. Haake, and M. Lewenstein,
Adiabatic drag and intial slip in random processes,
{\em Phys. Rev. A} {\bf 28} (1983) 3060--3612.

\bibitem{Hak78}
H. Haken,
{\em Synergetics, An introduction\/}
(Springer, 1978, second edition).

\bibitem{Kam85}
N.G. van Kampen,
Elimination of fast variables,
{\em Phys.\ Rep.\ }{\bf 124} (1985) 69--160.

\bibitem{Lei80}
C.E. Leith,
Nonlinear normal mode initialisation and quasi-geostrophy theory,
{\em J.\ Atmos.\ Sci.\ }{\bf 37} (1980) 958--968.

\bibitem{Lor86}
E.N. Lorenz,
On the existence of a slow manifold,
{\em J.\ Atmos.\ Sci.\ }{\bf 43} (1986) 1547--1557.

\bibitem{MR90}
G.N. Mercer and A.J. Roberts,
A centre manifold description of contaminant dispersion in channels
with varying flow properties,
{\em SIAM J.  Appl.\ Math.\ }{\bf 50} (1990) 1547--1565.

\bibitem{MR94}
G.N. Mercer and A.J. Roberts,
A complete model of shear dispersion in pipes,
{\em Jap.\ J. Ind.\ Appl.\ Math.\ }(1994) to appear.

\bibitem{MP86}
E. Meron and  I. Procacia,
Theory of chaos in surface waves: the reduction from hydrodynamics to
few-dimensional dynamics,
{\em Phys.\ Rev.\ Lett.\ }{\bf 56} (1986) 1323--1326.

\bibitem{Miel88}
A. Mielke,
On Saint-Venant's problem for an elastic strip,
{\em Proc.\ Roy.\ Soc.\ Edin.\ A} {\bf 110} (1988) 161--181.

\bibitem{Rob85}
A.J. Roberts,
Simple examples of the derivation of amplitude equations for systems
of equations possessing bifurcations,
{\em J.\ Austral.\ Maths.\ Soc.\ Ser.\ B} {\bf 27} (1985) 48--65.

\bibitem{Rob88}
A.J. Roberts,
The application of centre-manifold theory to systems which vary
slowly in space,
{\em J.  Austral.\ Math.\ Soc.\ Ser.\ B} {\bf 29} (1988) 480--500.

\bibitem{Rob89a}
A.J. Roberts,
The utility of an invariant manifold description of the evolution
of a dynamical system,
{\em SIAM J.\ Math.\ Anal.\ }{\bf 20} (1989) 1447--1458.

\bibitem{Rob89b}
A.J. Roberts,
Appropriate initial conditions for asymptotic descriptions of the
long-term evolution of dynamical systems,
{\em J.\ Austral.\ Maths.\ Soc.\ B} {\bf 31} (1989) 48--75.

\bibitem{Rob92}
A.J. Roberts,
Planform evolution in convection---an embedded centre manifold,
{\em J.\ Austral.\ Maths.\ Soc.\ B} {\bf 34} (1992) 174--198.

\bibitem{Rob93}
A.J. Roberts,
The invariant manifold of beam deformations. Part 1: the simple circular rod,
{\em J. Elasticity } {\bf 30} (1993) 1--54.

\bibitem{Sij85}
J. Sijbrand,
Properties of center manifolds,
{\em Trans.\ Amer.\ Math.\ Soc.\ }{\bf 289} (1985) 431--469.

\bibitem{Tay53}
G.I. Taylor,
Dispersion of soluble matter in solvent flowing slowly through a tube,
{\em Proc.\ Roy.\ Soc.\ Lond.\ }{\bf A 219} (1953) 186--203.

\bibitem{T80}
U.M.~Titulaer,
Corrections to the Smoluchowski equation in the presence of
hydrodynamic interactions,
{\em Physica A} {\bf 100} (1980) 251--265.

\bibitem{WR94}
S.D. Watt and A.J. Roberts,
The accurate dynamic modelling of contaminant dispersion in channels,
{\em SIAM J. Appl.\ Math.\ }(1994) to appear.

\bibitem{Win74}
A. Winfree,
Patterns of phase compromise in biological cycles,
{\em J.\ Math.\ Biol.\ }{\bf 1} (1974) 73--95.

\bibitem{WB87}
D. Wycoff  and N.L. Balasz,
Separation of fast and slow variables for a linear system by the
method of multiple time scales,
{\em Physica A} {\bf 146} (1987) 219--241.

\end{thebibliography}
\end{document}